\documentstyle[prl,aps,multicol,prb]{revtex}
\renewcommand{\narrowtext}{\begin{multicols}{2} \global\columnwidth20.5pc}
\renewcommand{\widetext}{\end{multicols} \global\columnwidth42.5pc}
\input{epsf.tex}

\begin{document}
\bibliographystyle{prsty}
\draft
%
\title{Giant oscillations of acoustoelectric current in a quantum channel}
\author{Harald Totland,$^{(1)}$ and Yuri Galperin,$^{(1,2)}$
}%
\address{
$^{(1)}$Department of Physics, University of Oslo, P.O.~Box 1048 Blindern,
N--0316 Oslo, Norway \\ 
$^{(2)}$ A. F. Ioffe Physico-Technical Institute RAS, 194021 St.
Petersburg, Russia 
}
\date{\today} \maketitle

\begin{abstract} 
A theory of d.c. electric current induced in a quantum channel by
a propagating surface acoustic wave (acoustoelectric current) is worked
out. The first observation of the acoustoelectric current in such a
situation was reported by J.~M.~Shilton et al., Journ.~Phys.~C (to be
published).  The authors observed a very specific behavior of the
acoustoelectric current in a quasi-one-dimensional channel defined in a
GaAs-AlGaAs heterostructure by a split-gate depletion
-- giant oscillations as a function of the gate voltage. Such a
behavior was qualitatively explained by an interplay between the
energy-momentum conservation law for the electrons in the upper
transverse mode with a finite temperature splitting of the Fermi
level. In the present paper, a more detailed theory is developed, and
important limiting cases are considered.   
\end{abstract}


\narrowtext

\section{Introduction}

During recent years much attention has been attracted by the interaction
between surface acoustic waves (SAW) and two-dimensional electron gases
(2DEG). Most experiments were performed in GaAs-Al$_x$Ga$_{1-x}$As
heterostructures, and the effects linear in acoustic intensity -- the sound
attenuation and the change of sound velocity -- 
were studied\cite{wkw,wsw,wpl,wrw}.
These
works were aimed to investigate the linear response of a 2DEG to a.c.\ strain
and electric fields without electric contacts attached to the
sample. In particular, very interesting experimental studies of these
effects in the quantum Hall regime were carried out in the
works\cite{wkw,wrw}.

The second class of studies deals with the so-called {\em
acoustoelectric effect}, namely a drag of 2D electrons by a traveling
SAW\cite{eww,smt,smr,eg,fmi}.
This effect is due to a transfer
of momentum from the SAW to the electrons due to SAW-electron
interaction.
 As a result, a d.c.\ {\em acoustoelectric current} appears
in  a closed circuit. If the circuit is open, the SAW generates a d.c.\
{\em acoustoelectric voltage} across the sample. In the simplest case, the
current (or the voltage) is proportional to the SAW {\em intensity}
rather than to its amplitude. It is a simple non-linear effect that
can be also employed to study the 2DEG.   

The first observation of an acoustoelectric current through a quantum
point contact was reported in\cite{tal}. The authors observed a
d.c.\ current in a quasi-one-dimensional channel defined in a
GaAs-AlGaAs heterostructure by a split-gate depletion. The conduction
of such channels is quantized -- its dependence
on the applied gate voltage consists of a set of plateaus divided by
sharp steps\cite{vW,wp}.  

An important point is that a very specific behavior of the
acoustoelectric current as a function of the applied gate voltage was
observed. Namely, 
the current is {\em not quantized}. On the contrary, it experiences
{\em giant oscillations} as a function of gate voltage, having maxima
near the steps between the plateaus of the conductance. 

A semi-quantitative explanation given in\cite{tal} attributes the
oscillations to the the electrons in the upper transverse sub-band of
the channel. When the bottom of this band is close to the Fermi
surface (i.e.\ near the step of conductance), these electrons at the
Fermi level have low longitudinal velocity. At some value of the gate
voltage this velocity  can  be close to the sound velocity $w$. Such
electrons interact strongly with the wave because they move synchronously to
the latter. At other gate voltages there are no electrons which are
able to interact effectively, and the drag is not efficient.

In the following, we will discuss the acoustoelectric effect
theoretically in more detail and consider the most important parameters
and limiting cases.

\section{Formulation of the problem}

Consider a ballistic point contact between two regions of a
2DEG. According to the Landauer formula, the
conductance  of such a junction is determined by transparencies $T_n$
which correspond to different transverse modes  
\begin{equation} \label{tm}
|nk\rangle =\chi_n (z,x)\exp\left(i\int^x k (\xi)\, d\xi \right)\, ,
\end{equation}
the functions $\chi_n$ and the wave-vector $k$ being slowly dependent on $x$.   
Here we label $k$ the wave vector at $x \rightarrow \infty$.
We have
\begin{equation} \label {fl}
G=\frac{2e^2}{\pi \hbar}\int_{\epsilon_n}^{\infty} \sum_{n=1}^N d\epsilon_{nk}
 \left(-\frac{\partial 
f_0(\epsilon_{nk})}{\partial \epsilon_{nk}}\right) \, T_n (\epsilon_{nk}) \, ,
\end{equation}
where $f_0(\epsilon_{nk})$ is the Fermi function, while
$\epsilon_{nk} = \epsilon_n + \hbar^2 k^2/2m$  is the electron energy
for the $n$th transverse mode. Such a concept is
based upon the assumption that electron thermalization takes place within
the region $\sim \ell_{\text{in}}$ (inelastic relaxation length) near
the contact. that allows one to reduce the problem to calculation of
transmittance of non-equilibrium electrons. 

One can imagine several sources of influence of the acoustic wave on
the current through the contact. The first one is a drag of the 2DEG in
the leads. Such a drag produces a ``phonon wind'' in the leads which
has been estimated by Kozub and Rudin\cite{kr} on the basis of the
model introduced by Streda\cite{str}. According to this model, one can
introduce the electron-phonon collision integral {\em in the leads},
${\hat I}^{\text{drag}}_{\text{e-ph}} (\epsilon)$, as a source of the force
acting upon the electrons. Then, the current through the contact is
\begin{equation} \label{pw}
I \propto \sum_n \int_{\epsilon_n}^{\infty} d\epsilon \, {\hat
I}^{\text{drag}}_{\text{e-ph}} (\epsilon_{nk}) \,
\left(-\frac{\partial
f_0(\epsilon_{nk})}{\partial \epsilon_{nk}}\right)
T_n(\epsilon_{nk}) \, ,
\end{equation}  
where  ${\hat
I}^{\text{drag}}_{\text{e-ph}} (\epsilon_{nk})$ is a smooth function,
determined by the properties of the leads. Consequently, such a
contribution has the form of a sum of steps similar to the conductance that is
explicitly stated in\cite{kr}. 

However, the piezoelectric interaction between the SAW and the electron
system in the leads is significantly screened by the 2DEG. Consequently,
the steps are very small comparing with the effects observed in the
experiment\cite{tal}, where maxima of the acoustoelectric drag current
rather than steps were observed.

As a result, to explain the observed experimental results we are left
with the region of the quantum channel where  the piezoelectric field
is not too much screened. Namely, we analyze the
drag due to a momentum transfer from the SAW to the electron gas {\em
inside the quantum channel}. The physical reason of the importance of
this region in comparison with the much larger region of the 2DEG is
the absence of effective screening. Such a physical picture implies
that one  has to consider SAW-electron interaction inside the channel
rather than employ B\"uttiker-Landauer formalism.

\section{General expressions for acoustoelectric current}

In the presence of a harmonic acoustic  wave with the frequency
$\omega$ and wave vector $\bf q$ the electrons acquire a perturbation
$${\cal H}_{\text{int}}=U \sum_{nk,n'k'}[C_{nk,n'k'}(q)
a^{\dag}_{nk}a_{n'k'} + {\text{h.c.}}]\, , $$
where
$$C_{nk,n'k'}(q)=\langle nk|\exp(iqx)|n'k' \rangle \, .$$ 
For simplicity, let us model the quantum point contact as
a channel with uniform width $d$. Such an approximation is valid if the
product of the channel's length $L$ times the SAW wave vector
$q$ is much greater than 1, $qL \gg 1$.\cite{end1} We also assume that
the inequality $qd \ll 1$ holds. The last assumption allows one to
neglect the inter-mode transitions due to SAW and to take into account
only diagonal in the mode quantum numbers $n,n'$ contribution. 

In this approximation,
$C=\beta_n (q)\delta_{nn'}\delta (k-k'-q) $, and  we can consider the
electrons of the $n$-th mode as moving in the effective classical field
$$V_n(x)=\Re [U\beta_n \exp(iqx-i\omega t)]\, .$$
Having in mind low enough frequencies,
$$\hbar q \ll mw, p_{\text F}\, ,$$
 one can use the Boltzmann equation for the
occupation number of the $n$-th 
mode, $f_{nk}(x)$ (see, e.g. Ref.~\onlinecite{gur}).
This equation has the form
$$\left(\frac{\partial}{\partial t} + v\frac{\partial}{\partial x} -
\frac{1}{m}\frac{\partial V_n}{\partial x}\frac{\partial}{\partial v}+
{\hat I}\right)f_{nk}(x)=0\, ,$$
where $ \hat I$ is the  operator describing relaxation of the
non-equilibrium distribution. We'll specify its form later, taking into
account two possible relaxation mechanisms: ({\em i}) impurity scattering,
and ({\em ii}) escape from the channel.

Let us expand $f_{nk}(x)$ in powers of the sound amplitude $U$: 
\begin{equation} \label{it}
f_{nk}(x)\equiv  f_0[\epsilon_{nk}+V_n(x)]+f_1+f_2 \, ,
\end{equation}
where
\begin{equation}
f_1 = \Re[f_{1\omega}\exp(iqx-i\omega t)]\, ;~~
f_{1\omega} \propto U\beta_n.
\end{equation}
The second-order part, $f_2 \propto  |U\beta_n|^2$,
is a sum of two items -- a stationary
part, and a part which varies in time with the frequency $2 \omega$.
We are interested only in the 1st part.
The second part does not contribute to d.c. current and will be omitted.
We get from (\ref{it}):
\begin{eqnarray} \label{it1}
{\hat B}f_{1\omega}&=&-i\omega U\beta_n\left(-\frac{\partial
f_0(\epsilon_{nk})}{\partial \epsilon_{nk}}\right) \, ,
\nonumber \\
{\hat I}f_2&=&\left< \frac{\partial V_n (x)}{\partial x}\frac{\partial
f_1}{\partial p}\right>_t\, .
\end{eqnarray}
Here we have introduced the operator 
$${\hat B}(q,\omega) \equiv {\hat I}+i(qv-\omega)$$
having the meaning of the operator of the linearized kinetic equation.
Angular brackets mean average over time $t$, $\langle \cdots \rangle_t
\equiv (\omega/2\pi)\int_0^{2\pi/\omega} \cdots dt $.
Substituting $f_1$ and using the relationship
$$ \langle\,\Re[C_\omega\exp(i\omega t)]\,
            \Re[D_\omega\exp(i\omega t)]\,\rangle_t
=(1/2)\Re (C_\omega D_\omega^*)$$
we obtain
\begin{equation}
f_2=-\frac{1}{2}|U \beta_n|^2q\omega
{\hat I}^{-1}\Re\left\{\frac{\partial}{\partial p}
\left[ {\hat B}^{-1}\left(-\frac{\partial
f_0(\epsilon_{nk})}{\partial \epsilon_{nk}}\right)\right] \right\}\, .
\end{equation}
As a result, we arrive at the following formal expression for
the acoustoelectric current (taking into account the sum over spins): 
\begin{eqnarray} \label{cur}
j&=&eU^2q\omega\sum_n |\beta_n|^2
\nonumber \\ && \times
\Re \int \frac{dp}{2\pi
\hbar}
v{\hat I}^{-1}\left\{\frac{\partial}{\partial p}
\left[ {\hat B}^{-1}\left(-\frac{\partial
f_0(\epsilon_{nk})}{\partial \epsilon_{nk}}\right)\right] \right\}\, .
\end{eqnarray}

\section{Analysis of important limiting cases} 

Eq.~(\ref{cur}) is the formal expression for the acoustoelectric
current. To evaluate it one needs to specify the relaxation operator
$\hat I$, i.e.\ to discuss sources of relaxation. The conventional way to
treat relaxation in quantum channel is to discuss scattering by
individual defects described by scattering matrices. However, we want
to emphasize that the problem of acoustoelectric effect has an
important specifics. Indeed, as we will demonstrate later, the effect
is due to electrons with small momenta $k$. For these selected
electrons the effective mean free path, $\ell_{\text{eff}}$, can be
much shorter than the mean elastic free path $\ell$ for a bulk
2DEG (see below). We believe that for realistic ballistic channels  the
inequalities $\ell > L > \ell_{\text{eff}}$ hold. As a result, one can
use a relaxation rate approach, similar to the conventional one for
bulk systems. The possibility for a non-equilibrium electron to escape
the channel will be allowed for in a model way.


\subsection{Relaxation rates}

Here we discuss two sources of relaxation,
 namely an elastic scattering within the channel (with the rate
$1/\tau$) and an escape from the channel due to its finite length $L$
(see Fig.~1), an effective rate being $|v_n|/L$.
\begin{figure}
\centerline{\epsfxsize=6cm \epsfbox{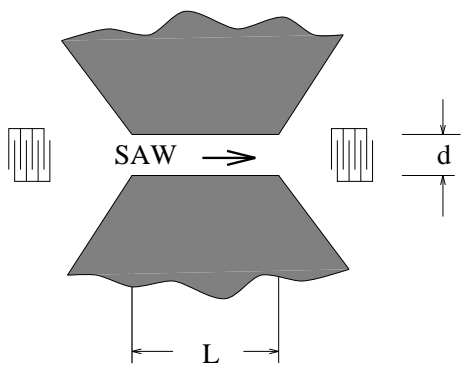}}

\bigskip
\caption{Schematic shape of QPC}
\end{figure}
\noindent 
 Here $v_n$ is
defined as the velocity for the $n$-th mode at a given energy $\epsilon$, 
$$\epsilon_{n,mv_n/\hbar}=\epsilon\, . $$
For lower transverse modes $v_n \sim v_{\text F}$, while for the upper
one (with the number $N$) it can be small. On the other hand, because
of low longitudinal 
electron velocity, the electrons of the upper mode are much more
effectively scattered by  impurities (see below).  
{}From this fact we come to an important conclusion:
for the upper mode (which is responsible for the acoustoelectric
effect near its maxima) the  impurity scattering might be important
even if the contact is ballistic. We would like to use this
opportunity to emphasize once more that in mesoscopic systems an interplay
between the impurity and other mechanisms of scattering is very much
dependent on the problem in question (cf.\ with\cite{jsg}).   

Introducing a correlation function of random impurity  
potential as
$${\cal K}(\bbox{r})=\langle V(\bbox{r})V(0)\rangle_{\text{im}}$$ 
and its matrix elements
$$K_{mk',nk} =\langle nk|{\cal K}(\bbox{r})|mk'\rangle \, , $$ we get   
\begin{equation} \label{rr}
\frac{1}{\tau_n (k)}=\frac{2\pi}{\hbar}\sum_{mk'}|K_{mk',nk}|^2 \delta
(\epsilon_{nk}-\epsilon_{mk'})\, .
\end{equation}
If the impurity potential has short range, then $|K_{mk',nk}|^2$ is
independent of the arguments $(mk',nk)$. Consequently,
$$\frac{1}{\tau_n(k)}=\frac{2\pi|K|^2}{\hbar}\nu (\epsilon_{nk})$$
where $\nu(\epsilon)$ is the density of states per given spin,
$$ \nu (\epsilon)=\frac{1}{\pi \hbar}\sum_{n=1}^N \frac {1}{|v_n|} \,
.$$
Assuming that the channel behaves as a rectangular box with the
thickness  $d$ in the transverse direction, we get $N \approx
\sqrt{\epsilon_{\text F}/\epsilon^* }$,
where $\epsilon^* = h^2/8md^2$.\cite{end2}

 For a wide contact ($N \rightarrow\infty$)
$\nu =\nu_bd$, where $\nu_b=m/2\pi\hbar^2$ is the 2D density of
states. As a result, in a channel we arrive at a smooth part of
the relaxation rate (which is of the order of the momentum relaxation
rate in the bulk 2DEG, $1/\tau_b$) plus an oscillating term 
$$\frac{2}{\pi\tau_b} \sqrt{\frac{\epsilon^*}{\epsilon - \epsilon_N}}
\, . $$
Finally, for all the levels except the highest one, one can put
$1/\tau_b$ for the impurity relaxation rate, while for the highest one
the appropriate estimate is
\begin{equation} \label{tim1}
\frac{2}{\pi\tau_b}\frac{v^*}{|v_N|}\, ,
\end{equation}
where $v^* = \pi \hbar/md$.
In high-mobility selectively-doped structures the scattering potential
is smooth. Consequently, the matrix elements $K_{nk',nk}$ significantly
increase with the decrease of $k$. As a result, impurity scattering
for the upper mode is much stronger than for the lower
ones\cite{timp}, and the expression (\ref{tim1}) acquires an
additional large factor $\alpha \sim
|K_{N,k_s;N,-k_s}/K_{k_F,k'_F}|^2$, where $k_s \sim mw/\hbar$.

To take account of the finite length of the contact we introduce also
the relaxation rate $|v_n|/L$. As a result,   
an estimate for the relaxation operator of the upper level is 
\begin{equation} \label{co}
{\hat I}=\max
\left( \frac{1}{\tau_b}\frac{v^*}{|v_N|}, \frac{|v_N|}{L}\right) \, .
\end{equation} 
We observe that there is a border value $v_c$ of $v_N$,
\begin{equation} \label{vc}
v_c=\max \left(v_{\text F}\sqrt{\frac{\alpha}{N} \frac{L}{\ell_b}}, 
v_{\text F} \right)  \, , 
\end{equation} 
where both mechanism give contributions of the same order of magnitude.
Here $\ell_b$ is the mean free path in a bulk 2DEG.
At $|v_N| \le v_c$ impurity relaxation becomes more important than
finite size of the contact.

Expression (\ref{co}) needs a more detailed discussion. Consider a
ballistic one-dimensional pipe where the particles are subjected to
a constant force $F$. As a result, they are accelerated as 
$v(t)=v_0+ (F/m)t, \quad x(t)=v_0t + (F/2m)t^2\, .$
At the time $t$, the distance between a given particle
and a particle started after time $\theta$ is $\delta x =\theta[v_0 +
(F/m)t]$.   
Consequently, the product of the velocity at the time $t$ and the
local density, $ v(t)/\delta x (t) = 1/\theta$ remains constant keeping
constant the current density inside the pipe.

The total current  is determined by the difference
between the contributions of the particles with opposite directions of
initial velocity. The difference is proportional to $(F/m)t \approx
FL/mv_0$. Such a result can be reproduced by the order of magnitude by
the assumption (\ref{co}).

\subsection{Estimates of the acoustoelectric effect}

Now we are ready to make estimates. For simplicity, we model the
relaxation operator by the following interpolation expression:
\begin{equation} \label{co1}
{\hat I}=\frac{1}{L}\frac{v_N^2+v_c^2}{|v_N|} \, .
\end{equation}
At $v_N \gg v_c$ it approximately describes escape from the channel,
while at  $v_F \gg v_N \gg v_c$ it is $L$-independent and equivalent
to the inverse life time for the electron with a given $k$ due to
elastic scattering.
We have,
$$v_N{\hat I}^{-1}=L\frac{v_N |v_N|}{v_N^2+v_c^2}\, .$$
Integrating Eq.~(\ref{cur}) by parts and taking into account that
$$\frac{\partial (v_N{\hat I}^{-1})}{\partial
p_N}=\frac{2L}{m}\frac{|v_N| v_c^2}{(v_N^2 +v_c^2)^2}$$
we arrive at the following expression for the oscillating part of the 
 acoustoelectric current:
\begin{eqnarray} \label{cur1}
j&=&\frac{LeU^2\omega}{\pi
\hbar }|\beta_N|^2
\int_{-\infty}^{\infty} dv_N \, F_\eta (v_N)\, F_T(v_N) \, ,
\nonumber \\
  F_\eta (v_N)&=&
\frac{\eta v_N^2v_c^2}{(v_N^2 +v_c^2)[v_N^2(v_N-w)^2+ \eta^2(v_N^2
+v_c^2)^2]} \, , 
\nonumber \\ 
 F_T(v_N)&=& \frac{1}{4k_{\text B}T}\,
\frac{1}{\cosh^2
[m(v_N^2-v_{N{\text F}}^2)/4k_{\text B}T]}
\, .
\end{eqnarray}
Here $\eta = 1/qL$ has the meaning of the ratio between the acoustic
wave length and the length of the contact, $w=\omega/q$ is the sound
velocity, while $v_{N{\text F}}=v_N(\epsilon_{\text F})$.

We observe that the expression (\ref{cur1}) consists of the product of
two functions. The function $F_T$ has sharp maxima at
$v_N=v_{N{\text F}}$ (see Fig.~2), the width being
$$\delta_T =\min \left(\frac{k_{\text B}T}{mv_N},
\sqrt{\frac{2k_{\text B}T}{m}}\right)\, .$$    
The properties of $F_\eta$ depend upon $\eta$
and the ratio $w/v_c$. Here we consider the limiting case
\begin{equation} \label{neq}
w/v_c \ll \sqrt{\eta}\, ,
\end{equation}
which is relevant to the present experimental situation. In this
case the impurity scattering dominates for the important group of the
electrons. If inequality (\ref{neq}) is 
met one can put $w=0$, and the integrand is symmetric. Then
\begin{eqnarray} \label{f1}
F_\eta (v_N)&=&\frac{1}{v_c^2}f\left(\frac{v_N}{v_c}\right) \, ,
\nonumber \\
f(x)&=& \frac{x^2}{1+x^2}\, \frac{\eta}{x^4+\eta^2(1+x^2)^2} \, .
\end{eqnarray}
Let us consider important limiting cases.

\subsubsection{Short waves, \protect{$\eta \ll 1$}}

At $\eta \ll 1$ $f(x)$ has a maximum at $x=\sqrt{\eta}$ with the peak
value 0.5 and the width $\sim \sqrt{\eta}$. The shape of the
oscillations of the acoustoelectric current depends upon the
relationship between $v_c\sqrt{\eta}$ and $\delta_T$. At
$v_c\sqrt{\eta} \ll \delta_T$ one can replace 
$$f(x) \rightarrow \frac{\pi \sqrt{\eta}}{2\sqrt{2}}
\delta(x-\sqrt{\eta}) \, .$$ 
As a result,
\begin{equation}  \label{cur2}
j=\frac{LeU^2\omega\sqrt{\eta}}{\sqrt{2}
\hbar v_c }|\beta_N|^2
F_T(v_c\sqrt{\eta}) \, .
\end{equation}
In this case, we have a peak near $v_{N{\text F}}=0$ (i.e.\ exactly at
the step), its shape being determined by the derivative of the Fermi
function, $F_T$. 

At $v_c\sqrt{\eta} \gg \delta_T$ the function $F_T$ behaves as 
$$\delta[(m/2)(v_N^2-v_{N{\text F}}^2)]=(1/mv_{N{\text
F}})\delta (v_N-v_{N{\text F}}) \, .$$
Consequently,   
\begin{equation}  \label{cur3}
j=\frac{2LeU^2\omega}{\pi \hbar v_{N{\text F}}mv_c^2 }
|\beta_N|^2 f\left(\frac{v_{N{\text F}}}{v_c}\right)
\, .
\end{equation}
Note that in both cases the current is independent of the channel's
length $L$. It is natural, because we consider the situation where the
intra-channel impurity scattering is the most important relaxation
mechanism. 

At $w/v_c \gg \sqrt{\eta}$ the escape of non-equilibrium
electrons is most important. In this limit 
one obtains $j \propto L$.\cite{gurpev} Consequently, measurements of the $L$-dependence of the
acoustoelectric current would help to discriminate between different
relaxation mechanisms.

\subsubsection{Long waves, \protect{$\eta \gg 1$}}

If the wave length of the SAW is greater than the channel length,
the function $f(x)$ can be approximated as 
\begin{equation} \label{f2}
f(x) = \eta^{-1} \frac{x^2}{(1+x^2)^3} \, .
\end{equation}
It has a maximum at $x=1/\sqrt{2}$ and width of the order 1.
Consequently, in dimensional variables the width is of the order of
$v_c$. It has to be compared to the difference  
$$\delta_v=|v_N-v_{N-1}| \approx
\sqrt{4N\epsilon^*/m}\approx v_{\text F}N^{-1/2}\, .$$
 If 
$$v_c \gg \delta_v\, , \quad {\text {or}} \quad L\gg \ell$$
the oscillations are not pronounced. On the contrary, at $L \ll \ell$
the oscillating part is pronounced. Again, its shape is determined by
the relationship between the widths of the functions $F_\eta$ and
$F_T$. At $v_c \gg \delta_T$ the result is given by Eq.~(\ref{cur3})
with $f(x)$ taken from (\ref{f2}). In the opposite limiting case,
$f(x)$ can be replaced by
$(\pi/16\eta)\delta(x-1/\sqrt{2})$. Consequently, the shape is
determined by the function $F_T$, like in Eq.~(\ref{cur2})

\section{Discussion}

Let us discuss qualitatively the  picture of the acoustoelectric
effect. The linear response of the electrons to the SAW with a given
wave vector $q$ is proportional to the effective ``interaction time''
$(qv-\omega)^{-1}$, during which an electron with the velocity $v$
moves in an almost constant field. At small $v$, or near the resonance
($v=w$), this time diverges, and relaxation becomes
important. In fact, the coupling is proportional to $\frac{\bar
I}{(qv-\omega)^2+{\bar I}^2}$. Consequently, to get an effective
coupling both the electron velocity and the scattering rate  have to
be small.
     
In a homogeneous 2D system, the most important relaxation mechanism is
disorder-induced 
scattering, the scattering rate being proportional to the 2D density of
states. Thus, the electron-SAW coupling for such a system is
determined by the product $q\ell$.
In a point contact relaxation differs from the case of homogeneous
2DEG due to following reasons.
\begin{itemize}
\item The contact has {\em finite length} $L$, the corresponding rate
being $\sim |v|/L$. This rate decreases as $v \rightarrow 0$.
\item Density of states in a QPC is an oscillating function of the
energy, it diverges at the thresholds corresponding to the filling of
new levels. Consequently, the disorder-induced relaxation rate for the
upper mode increases at the threshold.    
\end{itemize}     
As a result, the total relaxation rate is a non-monotonous function of
the electron velocity. The analysis given above leads to the
conclusion that coupling is optimal for the electrons having their
velocities in a relatively narrow range, the central velocity $s \sim
v_c \sqrt{\eta}$ being
small comparing to the Fermi velocity . Above we have estimated the
position of the center and an 
effective width of the important region which is shown in Fig.~2.
\begin{figure}
\centerline{\epsfxsize=7cm \epsfbox{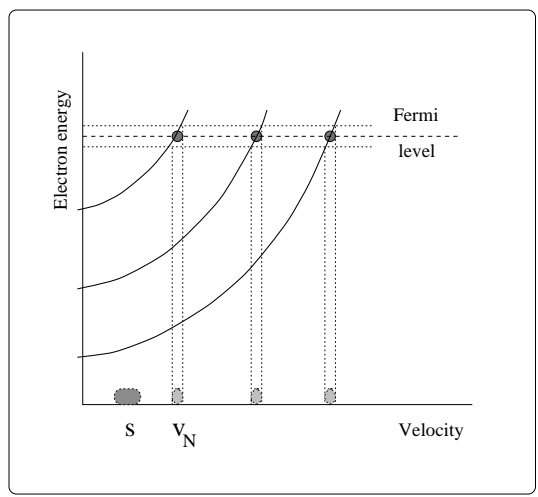}}

\bigskip
\caption{Scheme of energy and momentum conservation laws for
electron-SAW interaction}
\end{figure}

On the other hand, electronic states can contribute to interaction
only if their energies are in the vicinity of the Fermi
level. Consequently, for a given mode number $n$ the electron
velocities have to belong to a narrow interval centered around some
velocity $v_{N{\text F}}$. One can move these intervals by changing
the gate voltage.

To get a non-vanishing contribution to the current,
one of these regions has to overlap with the region centered
around $s$ (see the Fig.~2), which is possible only for the upper
level. As a result, the acoustoelectric current experiences {\em giant
oscillations} as a function of the gate voltage.
\cite{end3}
 Indeed, in the plateau region for the conductance 
the velocity $v_{N{\text F}}$ is large
enough and cannot overlap with the hatched region near $s$. Then, 
as the system is driven to the step, $v_{N{\text F}}$ decreases and
important regions start to overlap. Consequently, the current
increases.      

The fine structure of the peaks needs more careful discussion. The current
theory can lead to quantitative results only when impurity
relaxation is more important than the escape from the channel. In the
opposite limiting case the model assumption (\ref{co}) can provide
only order-of-magnitude estimates. The exact results in such a
situation depend both on the length and the shape of the channel\cite{gurpev}.

An important  source of the fine structure of the peaks  could be a 
gate-voltage-dependent screening of the 
coupling. This problems needs much more careful considerations. For a
rough estimate, we can introduce a screening factor as
$$U^2=\frac{U_0^2}{1+g[\sigma (q,\omega)/s]^2} \, , $$
where $\sigma (q,\omega)$ is the effective conductance while $g$ is a
(small) geometry-dependent dimensionless factor. The effective conductance can
be  estimated as a sum of the contributions of lower transverse modes
(which is a smooth function of the gate voltage) and of the contribution of the
upper mode
$$\sigma = \sigma_{\text {sm}} + \sigma_{\text {osc}}\, .$$
The latter can be estimated near the peak of the drag
current in the same way as for a long wire because usually $L/s \ge
\tau$. Considering a piece of wire with the length $\approx q^{-1}$,
we get the estimate  $C \sim q^{-1}$ for its capacitance (up to
logarithmic terms), and $R \sim q\sigma_1(q,\omega)$ for its
resistance. Here $\sigma_1$ is the 1D conductance calculated from the
kinetic equation. As a result, the dimensionless screening parameter
can be estimated as $(\omega RC)^{-1} \approx
q\sigma_1/s$. Consequently, the screening parameter is 
\begin{eqnarray} \label{ec}
\sigma_{\text{osc}} (q,\omega)/s & =& \frac{qe^2}{s\pi \hbar}
\int dp \, \left(v{\hat 
B}^{-1}v\right) \left(-\frac{\partial 
f_0(\epsilon_{nk})}{\partial \epsilon_{nk}}\right)
\nonumber \\
&&\sim \frac{e^2}{s\hbar}\, \frac{1}{qv_n\tau}
\, .
\end{eqnarray}
Far from the maximum of the acoustoelectric current this quantity 
appears small. However, near the maximum it can be of the order 1,
leading to a decrease of the current. Probably  it is the 
origin of the double-peak structure of the 1st peak of the
acoustoelectric current, observed in\cite{tal}. With the increase of
the number of occupied modes the 
smooth part of screening increases. As a result, the current
oscillations' amplitude must decrease, and the double-peak structure
has to be less pronounced. Such a behavior is in a qualitative
agreement with the experiment\cite{tal}. 

Unfortunately, it is very difficult to give realistic estimates for
the coupling constant connecting the intensity $S$ of SAW and the
amplitude $U$ of the electron's potential energy.
According to the experimental
results\cite{tal} we believe that it is determined by piezoelectric
interaction in the channel. Otherwise, under the conditions realized
in\cite{tal} steps predicted in\cite{kr} would be observed rather than
giant oscillations. However, the SAW in a layered structure has a
complicated polarization structure, and only a rough estimate can be
given
$$U_0 \sim \sqrt{\chi Se^2/\omega}\, . $$
Here $\chi$ is piezoelectric coupling constant which can be determined
by experiments on the change of sound velocity due to 2DEG. To check
the presented theory quantitatively it seems important to measure
acoustic intensity independently. 

\acknowledgements

We are grateful to V. I. Talyanskii for drawing
attention to the experimental results\cite{tal} and helpful
discussions, and to V. L. Gurevich for reading the manuscript and
important comments.  


\widetext

\begin{references}
\bibitem{end1} In the experiment{\protect{\cite{tal}}}, the point
contact had a shape of a quasi-one-dimensional channel with a length of
about 5000 {\AA}. The SAW had a frequency about 900 MHz, so the
product $qL$ was of the order 1. However,
we restrict ourselves with the limiting case $qL \gg 1$, where the
results are more explicit. We believe that at $qL \sim 1$ the general
behavior of the acoustoelectric effect will be the same.
\bibitem{end2}Strictly speaking, this estimate, as well as the following
discussions, are valid at $N \gg 1$. However, one can hope that the
estimates below are correct also for $N \ge 1$.
\end{references}
\end{document}